\documentclass[aps,12pt]{revtex4}
\usepackage{color}
\usepackage{epsfig}
\usepackage{amsmath}
\usepackage{graphicx}
\usepackage{subfigure}
\usepackage{ulem}
\usepackage{amsfonts}
\usepackage{epsfig,bm}
\usepackage{graphicx}
\usepackage{comment}
\usepackage{sidecap}
\usepackage{hyperref}
\newcommand{\be}{\begin{equation}}
\newcommand{\ee}{\end{equation}}
\newcommand{\ba}{\begin{eqnarray}}
\newcommand{\ea}{\end{eqnarray}}
\newcommand{\nn}{\nonumber}

\begin{document}

\title{Power Law Corrections to BTZ Black Hole Entropy}
\vspace{0.50cm}

\author{Dharm Veer Singh}
\email{veerdsingh@gmail.com}

\affiliation{Department of Physics, Centre of Advanced Studies,\\
 Banaras Hindu University, Varanasi - 221005, India}

\noindent

\begin{abstract}
\begin{center}
ABSTRACT
\end{center}
We study the quantum scalar field in the background of BTZ black hole and evaluate the entanglement entropy of the non-vacuum states. The entropy is proportional to the area of event horizon for the ground state, but the area law is violated in the case of non-vacuum states (first excited state and mixed states) and the corrections scale as power law.\vspace{5mm}\\
{keywords: Scalar Fields in BTZ Black Hole spacetime; Entanglement Entropy; Power Law Corrections.}
\end{abstract}

\maketitle

\setcounter{page}{0}

\newpage
\section{Introduction}

Black holes are singular solution of Einstein fields equation and are characterized by three parameter mass, charge and angular momentum classically (No hair theorem). Black holes possess the properties similar to a thermodynamic system and they can be assigned entropy and temperature. The black hole entropy also known as Bekenstein Hawking entropy behaves like the thermodynamic entropy and is proportional to area of event horizon of the black hole \cite{JD,JD1,JD2,JD3,JD4}. Hawking showed that the black hole emit radiation like a black body at finite temperature (Hawking radiation) and fixed the proportionality constant \cite{SWH}. The temperature of the black hole is identified with the surface gravity of the horizon. 
 
The origin of black hole entropy has remained mysterious for a long time. There are various approaches to understand the black hole entropy. In the simplest approach, one considers the quantum fields in the black hole spacetime and compute the black hole entropy by using entanglement entropy approach \cite{Bombelli,Srednicki}. The earliest attempt to understand the area law in terms of microscopic degrees of freedom refers to holographic principle by t'Hooft and Susskind \cite{TH,LS, JMM}. The direct counting of black hole microstates for BTZ black hole by Strominger and Vafa established the area law of black hole thermodynamics \cite{AS}. The AdS/CFT has provided further insights into the black hole entropy \cite{TT1,RMW1,Carlip,solo}.

In order to understand the statistical origin of black hole entropy in term of state counting, one has to understand the microscopic degrees of freedom of the black hole and the logarithmic of the number gives the Bekenstein Hawking entropy formula \cite{Cadoni:2009tk,Cadoni:2007vf}. In addition to this there are quantum corrections to black hole entropy known as logarithmic corrections and they arise due to the quantum fluctuations of the fields near the black hole horizon. An alternative way to understand the black hole entropy is provided by entanglement entropy of quantum fields in black hole spacetime.  Bombelli et al  and Srednicki have showed that this entropy arises due to the vacuum fluctuations entangled on the boundary surface and obtained by the averaging the quantum state of the complete system over the states of the fields located outside the boundary. 

The entanglement entropy appears to have different origin from the statistical entropy, but the apparent contradictions is resolved in semiclassical limit and the entanglement entropy shows agreement with Bekenstein-Hawking entropy (thermodynamics entropy). Thus, both the statistical entropy and entanglement entropy have thermodynamic limit, but this limit is achieved in a different manner in both cases \cite{Mann:1996ze,Cadoni:2010vf}. The ensemble average provides the necessary limit for the statistical entropy, but the entanglement entropy is obtained by averaging over the quantum correlations in the semiclassical limit.  

Here, we are interested in the first principle approach to BTZ black hole entropy in terms of entanglement of quantum fields in black hole spacetime \cite{Kim,Frolov,Carlip1}. The entanglement entropy provides the necessary information about the system inside and outside of the horizon and established a correlation between them. The behavior of quantum fields in background of the black hole (gravitational system) is not universal, but all of them show that the entropy obeys the area law for the ground state \cite{Casini,SM1}. 

The entanglement entropy measures the correlation between two subsystems, depending upon the geometry of the surface. The entanglement entropy calculation was first done by Bombelli et al \cite{Bombelli} and Srednicki \cite{Srednicki}, applied to a sphere. The leading term of entropy is $(\frac{R}{a})^2$, where ``$a$'' is the lattice spacing and ``$R$'' is the radius of the sphere. The coefficient of proportionality was calculated by these authors and it is not universal. Using this formalism Das et al \cite{SS,SS1,SS2,SS3,SS4} calculated the entropy of scalar field in the background of Schwarzschild black hole for the first excited state and mixed state. We consider the density matrix and calculate the entropy of the scalar field propagating in BTZ black hole spacetime by taking a partition close to horizon as measured by an observer at a proper distance of the order of cut off length. The boundary surface can be treated as the horizon of the black hole when scalar fields falls into the black hole horizon. 

In this paper, we study the corrections to the area law of the scalar fields in the background of BTZ black hole for first excited state and mixed states (superposition of ground state and first excited state). In ground state the entropy follows area law \cite{DS,DS1}, but for excited state the area-entropy relation is given by $S=a_1(\frac{r_+}{a})^{\mu}$ \cite{SS1}, where $r_+$ is the horizon radius. The value of exponent is less than unity $(\mu<1)$ in the case of excited state. This suggests that excited state violates the area law. In the mixed state, the wave function is linear superposition of ground state and excited state ($\Psi=a_3\Psi_0+a_4\Psi_1$). The entropy of mixed state is given by the relation $S_{MS}=b_3A\Big(1+\frac{b_4}{A^{\lambda}}\Big))$, where $b_3$ , $b_4$ and $\lambda$ are constants. In the large horizon limit, the correction term fall rapidly and  we recover the area law.

The paper is organized as follows. The formalism of quantum scalar fields in BTZ spacetime and the entanglement entropy of the system in ground state, first excited state and mixed state are reviewed briefly in section (\ref{EESBTZ}). The quantum scalar field in the background of BTZ black hole is studied in section (\ref{SFBTZ}). The numerical computation of entanglement entropy of  first excited state and mixed state is done in section (\ref{NCEE}). Finally, we discuss the results and its implications for black hole entropy in the section (\ref{RES}). 
\section{A Model Entanglement Entropy of Scalar field}
\label{EESBTZ}
In this section, we have  discussed the formalism of entanglement entropy of scalar field in BTZ background for ground state, first excited state and the mixed state,

\subsection{Ground State}

\label{EESBTZ1}
Let us consider a system of coupled harmonic oscillators $q^A$ $(A=1,.......,N)$ to study the entanglement entropy of the system, the Hamiltonian of the system is written as,

\be
H=\frac{1}{2a}\delta^{AB}p_Ap_B+\frac{1}{2}V_{AB}q^Aq^B,
\ee
where canonical momentum corresponding to the $q^A$ is $ p_A=a\delta_{AB}\,\dot{q}^B$, $\delta_{AB}$ is Kronecker delta and $V_{AB}$ is real, symmetric, positive definite matrix and ``$a$'' is fundamental length characterizing the system.

The total Hamiltonian of the system is re-written as,

\be
H=\frac{1}{2a}\delta^{AB}(p_A+iW_{AC}q^C)(p_B-iW_{BD}q^D)+\frac{1}{2a}\,Tr W
\ee
where $W$ is symmetric, positive definite matrix satisfying the condition $V_{AB}=W_{AC}W^C_B$. $(p_{A}+iW_{AC}q^{C})$ and $(p_{B}-iW_{BD}q^{D})$ are the creation and annihilation operators similar to harmonic oscillator problem and they obey the commutation relation,

\be
[a_A,a_B^{\dagger}]=2W_{AB}.
\ee
If $\psi_{0}$ is the ground state for the harmonic oscillator system, thus it follows the condition $(p_{A}-iW_{AC}q^{C})|\psi_{GS}>=0$ and the solution is given by, \cite{Bombelli}

\ba
\psi_{GS}(\{q^C\})&=&<\{q^C\}|\psi_{GS}>\nn\\
&=&\Big[\det\frac{W}{\pi}\Big]^{1/4} exp (-\frac{1}{2}W_{AB}\,q^{A}\,q^{B}).
\ea
Here ${q^{A}}$ splits into two subsystems, ${\{q^{a}\}}$ $(a=1,2,.......n_B)$ and $\{q^\alpha\}$ $(\alpha=n_{B+1},n_{B+2},.......N)$\footnotetext[1]{The subsystem $\{q^a\}$ and subsystem $\{q^{\alpha}\} $ regards as the inside and outside mode of the horizon.}. The matrix ``$W$'' naturally splits into four blocks as,

\[ (W)_{AB} = \left(\begin{array}{ccc}
A_{ab} & B_{a\beta} \\
B^T_{\alpha b} & D_{\alpha \beta}\end{array} \right).\]

The reduced density matrix of the subsystem ``$1$'' is obtained by tracing the degrees of freedom of the subsystem ``$2$''\footnotetext[2]{The subsystem ``$1$'' and subsystem ``$2$'' refers to the former subsystem ``$a$'' and later subsystem ``$\alpha$''. }, and is given by;

\ba
\rho_{red}(\{q^{a}\},\{q^{\prime b}\})&=&\int \Pi\, dq^c\langle\{q^{a},q^{\alpha}\}|\rho_{0}|\{q^{\prime b},q^{\beta}\}\nn\\
&=&(det\frac{M}{\pi})^{1/2}exp[-\frac{1}{2}M_{ab}(q^{a}q^{b} +q'^{a}q'^{b})] [-\frac{1}{4}(N)_{ab}(q-q')^a (q-q')^b]\nn\\
\label{rdm}
\ea
where,
\ba
&&M_{ab}=(A-BD^{-1}B^{T})_{ab}\nn\\
&&N_{ab}=(B^TA^{-1}B)_{ab}.
\ea
The reduced density matrix of the system ``$1$'' is obtained by tracing the degrees of freedom of the system ``$2$'' and is same as above equation (\ref{rdm}).

The system can be diagonalized by the unitary matrix $U$ and the transformations
$q^{a}\rightarrow \tilde{q}^a=(UM^{1/2})^a_bq^{b}$. Thus the density matrix reduces to \cite{Bombelli}, 

\be
\rho_{red}(\{q^{a}\},\{q^{b}\})=\Pi_{n}\Big[\pi^{-1/2}\exp\Big(-\frac{1}{2}(q_n q^n+q^{\prime}_n q^{\prime n}-\frac{1}{4}\lambda_i (q-q^{\prime})_n (q-q^{\prime})^n)\Big)\Big],
\ee
where $\lambda_i$ are the eigenvalues of the matrix $\Lambda^a_b=(M^{-1})^{ac}N_{cb}$. 

The entropy $S_{ent}=-Tr(\rho_{red}~\ln\rho_{red})$ of the system is given by,

\be
 S=-\sum_i \ln(\frac{1}{2}\lambda_i^{1/2})+(1+\lambda_i)^{1/2}\ln[(1+\lambda_i^{-1}+\lambda_i^{-1/2}],
\ee
\\
The entanglement entropy of the system is given by the relation,

\ba
S=\sum_{i=1}^{N-n_{B}}S_{i}\nn
\ea
where

\be
S_{i}=-\frac{\mu_i}{1-\mu_i}ln\,\mu_i-ln\,(1-\mu_{i})
\label{eqn1}
\ee
where $\mu_i:=\lambda_{i}^{-1}(\sqrt{1+\lambda_{i}}-1)^2$, $0<\mu_i<1$.
\subsection{First Excited State}
If $\psi_{GS}$ is the ground state for the harmonic oscillator system, then the first excited state $\psi_{ES}$ is calculated by the relation,

\ba 
\psi_{ES}(\{q^C\})&=&\gamma a^{\dagger}\psi_{GS}=\gamma (p_{A}+iW_{AC}q^{C})\Big[\det\frac{W}{\pi}\Big]^{1/4} exp (-\frac{1}{2}W_{AB}\,q^{A}\,q^{B})\nn\\
&=&\sqrt{2}\Big(\gamma^T\,W^{1/2}\delta^{AB}q_B\,\Big)\,\sum_{i=1}^{N}\,\Big(det\frac{W}{\pi}\Big)^{1/4}\exp\Big (-\frac{1}{2}W_{AB}\,q^{A}q^{B}\Big),\nn\\
&=&\sqrt{2}\Big(\gamma^T\,W^{1/2}\delta^{AB}q_B\Big)\,\psi_{GS}(\{q^C\}).
\ea

where $\gamma^T=\frac{1}{\sqrt{o}} $ (0,0,\ldots 0;1,1,\ldots 1), where ``$o$'' is the number of excitations and non-zero value of the row matrix, $\gamma^T\,\gamma=1$. The corresponding density matrix $(\rho=|\psi><\psi|)$ is given as,

\ba
\rho_{ES}\,(\{q^A\},\{q^{ \prime B}\}))&=&\int\prod_{i=1}^N\,dq^c\,\psi_{ES}(q^{A},q^{\prime B})\,\psi_{ES}^{\star}(q^{A},q^{\prime B})\nn\\
&=&\int\prod_{i=1}^N\,dq^c[q^{ {A}}\,W_{AB}^{\prime} q^{\prime B}]\,\psi_{GS}(q^{A},q^{\prime B})\,\psi_{GS}^{\star}(q^{A},q^{\prime B})
\ea
here ${q^{A}}$  again splits into two subsystems, ${{q^{a}}}$ $(a=1,2,.......n_B)$ and ${q^\alpha}$ $(\alpha=n_{B+1},n_{B+2},\ldots N)$. It is given by the relation $W^{\prime}=U^TW^{1/2}\,\gamma\,\gamma^TW^{1/2}U$. Then the $W^{\prime}$ decompose in the matrix form;

\[ W^{\prime}_{AB}= \frac{1}{2}yy^T=\left(\begin{array}{ccc}
{E_{ab}} & {F_{a\beta}} \\
{F}^T_{\alpha b}& H_{\alpha \beta} \end{array} \right),\]
where $E_{ab}=\frac{1}{2}y_Xy^T_X,~~~~F_{a\beta}=\frac{1}{2}y_Xy_Y^T,~~~~H_{\alpha\beta}=\frac{1}{2}y_Yy_Y^T$, and ``$y$'' is $N$ dimensional column vector and is given by, 

\[ y = \sqrt{2}U^TW^{1/2}\gamma=\left(\begin{array}{ccc}
y_{A} \\
y_{B}\end{array} \right).\]

The density matrix of the subsystem is ``$1$'' is obtained by tracing the degree of freedom of the subsystem ``$2$'' (``$1$'' and ``$2$'' refers to the subsystem ``$a$'' and ``$\alpha$'' respectively) and is given by,

\ba
\rho_{ES}\,(\{q^a\},\{q^ {\prime b}\}))&=&\int\Pi dq^c<\{q^a, q^{\alpha}\} |\rho_{GS}|\{q^{\prime b}, q^{\beta}\}>\nn\\
&=&\kappa\Big[\rho_{GS}\,(\{q^a\},\{q^ {\prime b}\}))\,\Big(-\frac{1}{2}G_{ab}{(q^{a}q^{b}+q^{\prime a}q^{{\prime \beta}})}\nn\\&&\quad\quad-\frac{1}{4}C_{ab}(q-q^{\prime})^{a}(q-q^{\prime})^{b}\Big)\,\rho_{GS}\,(\{q^{ a}\},\{q ^{\prime {b}}\})\Big]
\label{d1}
\ea
where,
 
\ba
&&C_{ab}=\frac{1}{\kappa}\Big(2E-FD^{-1}B^T-BD^{-1}F^T+BD^{-1}HD^{-1} B^T\Big)_{ab},\nn\\
&&G_{ab}=\frac{1}{\kappa}\Big(2FD^{-1}B^T-BD^{-1}HD^{-1} B^T\Big)_{ab}.\nn
\ea
This is density matrix of excited state and it is not similar to the ground state density matrix. So this matrix cannot be factorized into harmonic oscillator density matrix because the excited state density matrix does not contain exponential terms. It cannot be written as a product of harmonic oscillator systems. To compare this density matrix  of excited state (\ref{d1}) with the ground state (\ref{ES1}), we take the limit \cite{SS2},

\ba
&&\epsilon_1=q^T_{max}\,C\,q_{max}\ll1,\nn\\
&&\epsilon_2=q^T_{max}\,G\,q_{max}\ll1.
\ea
where 

\be
q^T_{max}=\frac{3(N-n)}{2 Tr(M_{ab}-N_{ab})}\,(1,1,\ldots).
\ee
In this limit, the density matrix can be written as,

\ba
&&1-\,\Big(\frac{1}{2}G_{a b}{(q^{\alpha}q^{b}+q^{\prime a}q^{{\prime b}})}+C_{a b}(q-q^{\prime})^{a}(q-q^{\prime})^{b}\Big)\nn\\&&\qquad\qquad\qquad\qquad=\exp\Big(-\frac{1}{2}G_{a
b}{(q^{ a}q^{b}+q^{\prime a}q^{{\prime b}})}-\frac{1}{4}C_{a b}(q-q^{\prime})^a(q-q^{\prime })^b\Big).\nn\\
\ea
Finally the reduced density matrix for the excited state is written as \cite{SS2},

\ba
\rho_{ES}\,(\{q^a\},\{q^ {\prime b}\}))&=&\kappa\exp\Big(-\frac{1}{2}G_{a b}^{\prime}{(q^{a}q^{b}+q^{\prime a}q^{{\prime b}})}-\frac{1}{4}C_{a b}^{\prime}(q-q^{\prime})^a(q-q^{\prime })^b\Big)\nn\\
\label{ES1}
\ea
where

\ba
&&(G^{\prime}_{a b}=M_{ab}+G_{a b}),\nn\\
&&(C^{\prime}_{a b}=N_{ab}+C_{a b}).
\ea
and
$ N_{a b}=(BD^{-1}B^T)_{a b}$, $ M_{a b}=(A-BD^{-1}B^T)_{a b}$, and $\kappa=Tr (EA^{-1})_{ab}$.

Now, this density matrix (\ref{ES1}) is similar to the ground state density matrix (\ref{rdm}). This can be factorized into the harmonic oscillator density matrix. The entropy of the excited state is computed using the procedure discussed in subsection (\ref{EESBTZ1}).

\subsection{Mixed State}

The mixed state is the linear superposition of ground state and first excited state, the wave function of the mixed state is given as,

\ba
\psi_{MS}(\{q^{A}\}) &=& c_0\,\psi_{GS} (\{q^{A}\})+c_1\psi_{ES}(\{q^{A}\})\nn\\
&=&\Big[c_0+c_1\,\sqrt{2}\gamma^T W_{AB}^{1/2}q^{A}q^{B}\,\Big]\,\psi_{GS}(\{q^{A}\})\nn\\
\ea
where $c_0$ and $c_1$ are the real constants with $c_o^2+c_1^2=1$.where $(c_0=1,c_1=0)$ for ground state, $(c_0=0,c_1=1)$ for excited state, $(c_0=c_1=\frac{1}{\sqrt{2}})$ for equal mixing and $(c_0=\frac{1}{2},c_1=\frac{\sqrt{3}}{2})$ for high mixing . 

The density matrix of the mixed state is written as,
\ba
\rho_{MS}(\{q^{A}\},\{q^{\prime B}\})&=&\int\prod_{i=1}^N\,dq^c\,\psi_{MS}(q^{A},q^{\prime B})\,\psi_{MS}^{\star}(q^{A},q^{\prime B})\nn\\
&=&c_0^2\,\rho_{GS}(\{q^{A}\},\{q^{\prime B}\})+c_1^2\,\rho_{ES}(\{q^{A}\},\{q^{\prime B}\})+c_0c_1\,\rho_{X}(\{q^{A}\},\{q^{\prime B}\}).\nn\\
\ea
where $\rho_X$ is the interaction term and it is written as,

\ba
\rho_X(\{q^{A}\},\{q^{\prime B}\})&=&\int\prod_{i=1}^n\,dq^c\Big[f(q^{A},q^{B})+f(q^{\prime A},q^{\prime B})\Big]\rho_{GS}(\{q^{A}\},\{q^{\prime B}\})\,\rho_{GS}^{\star}(\{q^{A}\},\{q^{\prime B}\}),\nn\\
&=&\Big(y_Y-BD^{-1}y_X\Big)(q+q^{\prime})^{A}\rho_{GS}(\{q^{A}\},\{q^{\prime B}\}).
\ea
Now again ${q^{A}}$  splits into two subsystems, ${{q^{a}}}$ $(a=1,2,.......n_B)$ and ${q^\alpha}$ $(\alpha=n_{B+1},n_{B+2},.......N)$. We can write the reduced density matrix of the system``$1$'' obtained by tracing the degree of freedom of the system ``$2$'', 

\ba
\rho_{MS}\,(\{q^a\},\{q^ {\prime b}\}))&=&\int\Pi dq^c<\{q^a, q^{\alpha}\} |(c_0\rho_{GS}+c_1\rho_{ES})|\{q^{\prime b}, q^{\beta}\}>\nn\\
&=&c_0^2+\Big[c_1^2\kappa_1+u(q^{a},q^{\prime b})+c_0c_1v(q^{a},q^{\prime b})\Big]\rho_{GS}(\{q^{a}\},\{q^{\prime b}\}),\nn\\
\ea
where $u(q^{a},q^{\prime b})=\Big(-\frac{1}{2}G_{a b}{(q^{a}q^{b}+q^{\prime a}{q^{\prime b}})}-\frac{1}{4}C_{a b}(q-q^{\prime a})(q-q^{\prime b})\Big)$,\\ and $v(q^{a},q^{\prime b})=(y_Y-p)^T(q+q^{\prime})^a$.

Now, we define the 

\be
 F(q^{a},q^{\prime b})=\Big[1+\kappa_1\,w(q^{a},q^{\prime b})+\kappa_2v(q^{a},q^{\prime b})+\frac{\kappa_2^2}{2}v^2(q^{a},q^{\prime b})\Big],
\ee

where  $k_1=\frac{c_1^2}{c_0^2+c_1^2k}$ and $k_2=\frac{c_0c_1}{c_0^2+c_1^2k}$. 

The density matrix of mixed state is re-written as,

\be
\rho_{MS}(\{q^{a}\},\{q^{\prime b}\})=\tilde{\kappa} F(q^{a},q^{\prime b})\rho_{GS}(\{q^{a}\},\{q^{\prime b}\}).
\ee
where

\ba
w(q^{a},q^{\prime b})=\Big(\frac{1}{2}G^{\prime\prime}_{a b}{(q^{a}q^{b}+q^{\prime \alpha}q^{{\prime b}})}+C^{\prime\prime}_{a b}(q-q^{\prime a})(q-q^{\prime b})\Big)
\ea

\ba 
&&G^{\prime\prime}_{ab}=\Big[G-2\kappa_0\Big(G-\frac{H}{\kappa}\Big)\Big]_{ab}, \nn\\ &&C^{\prime\prime}_{ab}=\Big[C+2\kappa_0\Big(G-\frac{H}{\kappa}\Big)\Big]_{ab}.
\ea
The value of $k_0=\frac{c_0^2}{c_0^2+c_1^2k}$, and $ {\tilde{\kappa}}={c_0^2+c_1^2k}$. 
We take the value of $c_0=c_1=\sqrt{\frac{1}{2}}$ for equal mixing and $c_0={\frac{1}{2}},~~ c_1=\frac{\sqrt{3}}{2}$ for high mixing.

This is the density matrix of mixed state and it is not similar to the ground state density matrix. So this matrix cannot be factorized into harmonic oscillator density matrix, because it does not contain the exponential terms. It can not be written as the product of harmonic oscillator systems. The limit can be taken as \cite{SS1},

\ba
&&\epsilon_1=q^T_{max}\,C^{\prime\prime}\,q_{max}\ll1,\nn\\ &&\epsilon_2=q^T_{max}\,G^{\prime\prime}\,q_{max}\ll1.
\ea
Then $F(q^{a},q^{\prime b})$ is re-defined as,

\be
F(q^{a},q^{\prime b})=\exp\Big({\tilde{\kappa_1}}w(q^{a},q^{\prime b})+{\tilde{\kappa_2}}v(q^{a},q^{\prime b})\Big)
\ee
Finally, the density matrix of the mixed is expressed in the form \cite{SS1},

\be
\rho_{MS}(\{q^{a}\}\{q^{\prime b}\}={\cal N}\exp\Big(-\frac{1}{2}G^{\prime\prime\prime}_{\alpha\beta}{(q^{a}q^{b}+q^{\prime a}q^{q^{\prime b}})}-\frac{1}{4}C^{\prime\prime\prime}_{ab}(q-q^{\prime a})(q-q^{\prime b})\Big)
\label{eqn:ms}
\ee
where

  \ba
&&C^{\prime\prime\prime}_{ab}=\Big[N+\kappa_1\,C^{\prime\prime}\Big]_{ab}=\Big[N+\kappa_1\,C-2\kappa_0\kappa_1\Big(G-\frac{E}{\kappa}\Big)\Big]_{ab},\nn\\
&&G^{\prime\prime\prime}_{ab}=\Big[M+\kappa_1\,G^{\prime\prime}\Big]_{ab}=\Big[M+\kappa_1\,G+2\kappa_0\kappa_1\Big(G-\frac{E}{\kappa}\Big)\Big]_{ab}.\nn
\ea
and ${\cal N}=-\kappa_2\Big(y_Y-BD^{-1}y_X\Big)$. 

The form of this density matrix (\ref{eqn:ms}) is similar to the ground state density matrix (\ref{rdm}) and can be used to calculate the entropy of the mixed state. Therefore the entropy of the mixed states can be calculated using the relation $S=-Tr(\rho\log\rho)$, by substituting the value of density matrix from equation (\ref{eqn:ms}) and then computing the associated entropy of the mixed states using the procedure as discussed in subsection (\ref{EESBTZ1}). 

\section{Scalar Fields in BTZ Black Hole spacetime}
\label{SFBTZ}
The action of the (2+1) dimensional gravity with cosmological constant can be considered as \cite{MB},

\be
S=\frac{1}{2\pi}\int d^3x\sqrt{-g}\,[R+2\Lambda],
\ee
where $\Lambda={-\frac{1}{l^2}}$ is the cosmological constant. The BTZ black hole is solution of (2+1) dimensional gravity with negative cosmological constant \cite{MB} and the metric is given by;

\be
{ds}^2=-(-M+\frac{{r}^2}{{l}^2}){dt}^2+\frac{1}{(-M+\frac{{r}^2}{{l}^2}+\frac{{J}^2}{{4r}^2})}{dr}^2+{r}^2{d\phi}^2-J{dt}{d\phi}~,
\ee
where $-\infty < t < \infty$ and $0\leq\phi\leq2\pi$. 

The solution is parameterized by the mass $M$ and angular momentum $J$ of the black hole and they obey, $M>0$ and $|J|< Ml$. The inner and outer horizon of the BTZ black hole are located at,

 \be 
 r_{\pm}=l\Big[\frac{M}{2}\Big( 1{\pm}\sqrt{1-(\frac{J}{Ml})^2}\Big )\Big]^{1/2}.
 \ee 
 In the external case $J=Ml$, the inner and outer horizons coincide.

The proper length from the horizon is given by $r^2=r_+^2\cosh^2\rho+r_-^2\sinh^2\rho$ and the metric of the black hole can be rewritten as;

\be
ds^2=-u^2dt^2+d\rho^2 +l^2\,(u^2+M)d\phi^2-Jdt\,d\phi,
\ee 
where, $r^2(\rho)=l^2(u^2+M)$.\\
The action of massless scalar field in the background of BTZ black hole is written as,

\be
S=-\frac{1}{2}\int dt\sqrt{-g}\,(g^{\mu\nu}(\partial_{\mu}\Phi\partial_{\nu}\Phi)).
\ee 
The conjugate momentum $\pi_m$ conjugate to $\Phi_m$, is given by,

\ba
\pi_m=\sqrt{\frac{(M+u^2 )}{(u^2+\frac{J^2}{4(u^2+M)})}},\dot{\Phi}_m+\frac{iJm}{2u\sqrt{[(u^2+M)+\frac{J^2}{4u^2}]}}\,{\Phi_m}\nn.
\ea
We define the new momentum to diagonalise the Hamiltonian is defined as,

\be
\tilde{\pi}_m=\pi_m-\frac{iJm}{u\sqrt{[(u^2+M)+\frac{J^2}{4u^2}]}}\Phi_m.
\ee
The canonical variable ($\phi_m, {\tilde \pi}_m$) satisfy the following relation,

\be
\{\phi_m(\rho), {\tilde \pi}_m(\rho)\}=\frac{iJm}{u\sqrt{[(u^2+M)+\frac{J^2}{4u^2}]}}\delta_{m,m^{\prime}}\delta(\rho-\rho^{\prime}).
\ee
Now, we define the variable $\psi$,

\ba
\psi_m(t,\rho)=\Big({\frac{{u^2+\frac{J^2}{4(u^2+M)}}}{{(M+u^2 )}}}\Big)^{1/4}\,\Phi_m(t,\rho),
\ea
The Hamiltonian of the scalar field in the BTZ background spacetime is \cite{DS},

\ba
H&=&\frac{1}{2}\int d\rho\,\tilde{\pi}_{m}^2(\rho)+\frac{1}{2}\int\,d\rho\, d\rho'\,u\sqrt{[(u^2+M)+\frac{J^2}{4u^2}]}\,\nn\\&&\qquad\qquad\Big(\partial_{\rho}(\sqrt{\frac{\sqrt{[u^2+\frac{J^2}{4(u^2+M)}]}}{\sqrt{(u^2+M)}}})\,\psi_m\Big)^2+m^2{\frac{{u^2+\frac{J^2}{4(u^2+M)}}}{{(M+u^2 )}}}\psi_m^2,
\ea
The system can be discretized as,

 \ba 
&& \rho\rightarrow(A-1/2)a,\nn\\
 &&\delta(\rho-\rho ')\rightarrow \frac{\delta_{AB}}{a}.
 \ea
where $A,B=1,2\ldots N$ and ``$a$'' is UV cut-off length. The continuum limit is regained by taking the $a\rightarrow 0$ and $N\rightarrow \infty$ while the size of the system remains fixed. The cut-off parameter, ``$a$''  is defined  as a  lattice spacing becomes the cut-off at Planck length in the quantum gravity.

The Hamiltonian of the system can be discretized using the replacements,

\ba
&&\psi_m(\rho)\rightarrow q^A,\nn\\
&& \tilde{\pi}_m(\rho)\rightarrow p_A/a,\nn\\
&& V(\rho,\rho^{\prime})\rightarrow V_{AB}/a^2.
\ea
and the corresponding discretized Hamiltonian of the scalar field in the BTZ black hole is identical to the the set of $N$ coupled harmonic oscillators. it is given as,

\be
H=\sum_{A,B=1}^{N}\Big[\frac{1}{2a}\delta^{AB}p_A\,p_B+\frac{1}{2}V_{AB}\,q_m^A\,q_m^B\Big]
\ee
where $p_{A}=a\delta_{AB}\dot{q}^{B}$ is the canonical momentum conjugate to $q^{A}$.

We discretized the Hamiltonian using the middle point prescription and replace the terms using the identity $f(\rho) \partial_{\rho}g(\rho)=\frac{1}{a}f_{A+1/2}(g_{A+1}-g_A)$,

\ba
V_{AB}^m q_m^A q_m^B&=&a\sum_{A=1}^N\Big[u_{A+\frac{1}{2}}\sqrt{(u^2_{A+\frac{1}{2}}+M)+\frac{J^2}{4u^2_{A+\frac{1}{2}}}}\Big(\sqrt{\frac{\sqrt{u^2_{A+1}+\frac{J^2}{4(u^2_{A+1}+M)}}}{\sqrt{u^2_{A+1}+M}}}\psi_m^{A+1}\nn\\&&\qquad\qquad-\sqrt{\frac{\sqrt{u^2_A+\frac{J^2}{4(u^2_A+M)}}}{\sqrt{u^2_A+m}}}\psi_m^{A}\Big)^2+\frac{m^2}{r_+}\,\frac{u^2_A+\frac{J^2}{4(u^2_A+M)}}{u^2_A+M}\psi_m^{A^2}\Big].
\label{mat}
\ea
here

\ba
&&u_A=\frac{u(\rho=(A-1/2)a)}{r_+},\nn\\&& u_{A+1/2}=\frac{u(\rho=Aa)}{r_+},\nn\\
&&\psi_m^A=\psi_m(\rho=(A-1/2)a).
\ea

The $(N \times N)$ matrix representation of the $V_{AB}^m$ is given by, 

\begin{eqnarray}
\left( V^{(m)}_{AB}\right) & = & \left( 
\begin{array}{cccccc}
\Sigma^{(m)}_1 & \Delta_1 & & & & \\
\Delta_1 & \Sigma^{(m)}_2 & \Delta_2 & & & \\
& \ddots & \ddots & \ddots & & \\
& & \Delta_{A-1} & \Sigma^{(m)}_A & \Delta_A & \\
& & & \ddots & \ddots & \ddots 
\end{array}
\right), \nonumber
\label{eqn:Vlm}
\end{eqnarray}

The matrix elements are,

\begin{eqnarray}
\Sigma_A^{(m)}&=&\frac{\sqrt{[u^2_{A}+\frac{J^2}{4(u^2_{A}+M)}]}}{\sqrt{(u^2_{A}+M)}}\Big(u_{A+1/2}\sqrt{[(u^2_{A+1/2}+M)+\frac{J^2}{4u^2_{A+1/2}}}\nonumber\\&&\qquad\qquad+\,u_{A-1/2}\sqrt{[(u^2_{A-1/2}+M)+\frac{J^2}{4u^2_{A-1/2}}}\Big)^2+m^2{\frac{{u^2_A+\frac{J^2}{4(u^2_A+M)}}}{{(M+u^2_A )}}},\nn\\
\end{eqnarray}

\begin{eqnarray}
\Delta_A&=&-u_{A+1/2}\sqrt{[(u^2_{A+1/2}+M)+\frac{J^2}{4u^2_{A+1/2}}}\nn\\&&\qquad\qquad\qquad\qquad\sqrt{\frac{\sqrt{[u^2_{A+1}+\frac{J^2}{4(u^2_{A+1}+M)}]}}{\sqrt{(u^2_{A+1}+M)}}}\sqrt{\frac{\sqrt{[u^2_{A}+\frac{J^2}{4(u^2_{A}+M)}]}}{\sqrt{(u^2_{A}+M)}}}.\nonumber\\
\end{eqnarray}
where $\Delta_A$ is a off diagonal term representing the interactions. This is compared with the corresponding Hamiltonian in the section (\ref{EESBTZ}) and evaluates the entanglement entropy of the first excited state and mixed state is evaluated.

The total entropy of the system is given as the sum of all modes ``$m$'' (azimuthal angular momentum) \cite{Huerta:2012},

\be
S=\lim_{N\rightarrow\infty}S(n_B,N)=S_0+2\sum_{m=1}^{\infty}S^{m}_{ent}.
\label{TE}
\ee
Where $S(n_B,N)$ is the entanglement entropy of the total system $N$ with partition size $n_B$. The $S_0$ is the entropy of the system at ($m=0$), and $S_{ent}^m$ is the entropy of the subsystem for given value azimuthal quantum number ``$m$''. The equation (\ref{TE}) is infinite series, and it converges at large value of ``$m$''. We truncate the series depending on the accuracy, which we required.
\section{Numerical Computation of Entanglement Entropy }
\label{NCEE}

In this section, we have computed the entropy of first excited state and mixed state using the procedure outlined in section (\ref{EESBTZ}). We plots the graph of entropy as a function of $R=n_B\,a$, where $n_B$ is the size of partition. The entropy varies linearly as a function of $(\frac{R}{a})$ in ground state entropy. In first excited state it is not a linear function of $(\frac{R}{a})$. The same result is obtained for the mixed states at small horizon radius, but for  the large horizon radius, it approaches to ground state .

\subsection {First Excited State}
In figure (\ref{tab:high1}), the entropy $S_{EH}$ as a function of $(\frac{r_+}{a})$ is plotted, where $r_+$ horizon radius of the BTZ black hole and ``$a$" is identified as fundamental scale of the system and can be taken as Planck length in the fundamental theory of gravity. From these plots it can interpreted that the entropy for excited state does not follow the area law and the relation gives the entropy of excited state is given by the relation,

\be
S=a_1(\frac{r_+}{a})^{\mu},
\ee
where ($\mu<1)$ )  the entropy does not follow the area law $(S=A/4)$. The $a_1$ and $\mu$ are the fitting parameter, which are calculated numerically. The numerical value of these parameter are tabulated in table (\ref{tab:high1}),

\begin{center}
\begin{table}[h]
\begin{center}
\begin{tabular}{|l|l|r|l|r|l|r|}
\hline
\multicolumn{1}{|c|}{ Parameters } & \multicolumn{1}{c|}{ $o=10$ } & \multicolumn{1}{c|}{$ o=30$ }& \multicolumn{1}{c|}{$o=50$} \\
\hline
\,\, $a_1$\,\, & \,\, 0.42\,\, &0.55\,\, & \,\, 0.75 
\\
\,\, $\mu$\,\, \,\, \,\, &\,\, 0.93\,\, & \,\,0.88\,\,& \,\, 0.81\,\, 
\\
\hline
\end{tabular}
\end{center}
\caption{For the fixed value of $N =200$, we calculate the entropy of the first excited state for the number of excitation $o=10,30,50$.}
\label{tab:high1}
\end{table}
\end{center} 
From above table the value of $\mu$ is always less then unity and the exponent $\mu$
 decreases with excitation number ``$o$'' and the slope ``$a_1$'' increases with the excitation number ``$o$''. It is impossible to recover the area law for excited state with increasing the excitation number ``$o$''. So we can interpret this as the slope of fitting lines are increasing with the number of excitations and decreasing the value of exponent tabulated in the table (\ref{tab:high1}). 
 
\begin{figure}
\centering
\includegraphics[width=.95\textwidth]{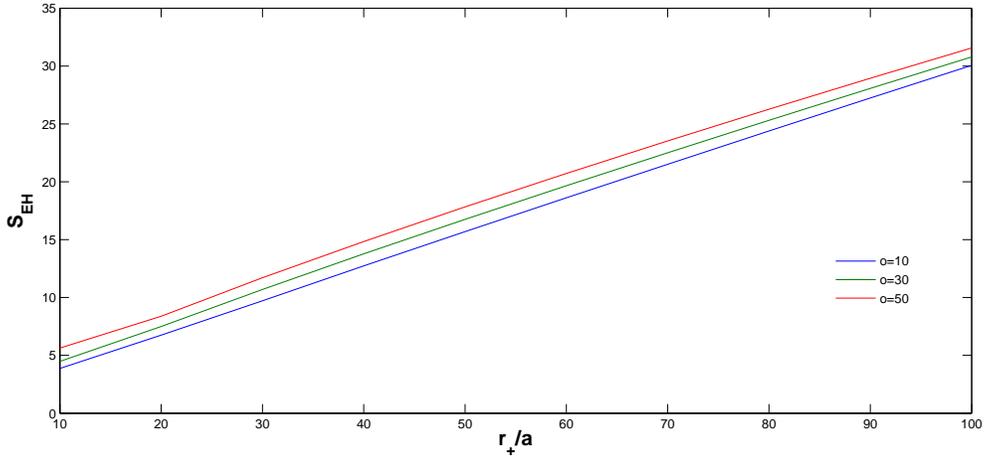}
\caption{Entanglement entropy for excited state $S_{ES}$ is shown as a functions of $r_+/a$ for $N=200$ with excitation $o=10$ (black), $o=30$ (blue) and $o=50$ (red).}
\label{figes}
\end{figure}
\subsection{Mixed States}

\begin{figure}
\centering
\includegraphics[width=1.05\textwidth]{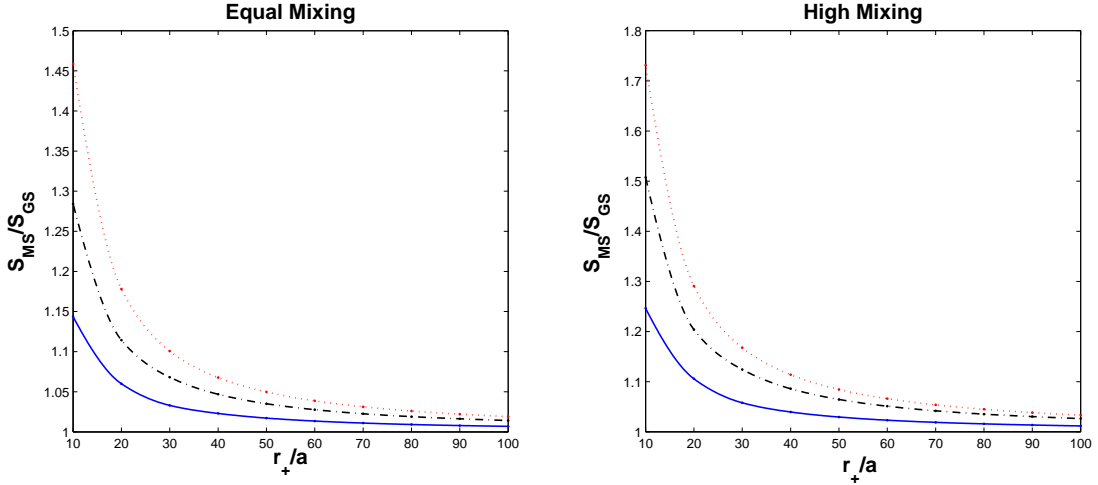}
\caption{Ratio of Entanglement entropy of mixed state for equal (high) mixing is shown as a functions of $r_+/a$ for $N=200$ with excitation $o$=10, 30 and 50.}
\label{figratio}
\end{figure}

The entropy of the mixed state is plotted in the figure (\ref{figratio}). This figure shows that  the entropy of mixed state approach  to the ground state entropy at large horizon radius. The relative entropy of the ground state vs. mixed state (for equal and high mixing) are shown in figure (\ref{figratio}) and the form of mixed state is writen as,

\be
\frac{S_{MS}}{S_{GS}}=b_1+{b_2}\Big(\frac{r_+}{a}\Big)^{-{\nu}}.
\label{MS}
\ee

Here, $b_1, b_2$ and $\nu$ are the fitting parameters. The value of parameters for equal and high mixing are tabulated in table (\ref{tab:high}), 

\begin{center}
\begin{table}[h]
\begin{center}
\begin{tabular}{|l|l|r|l|r|l|r|}
\hline
\multicolumn{1}{|c|}{ }&\multicolumn{1}{c}{ }&\multicolumn{1}{c}{ Equal }&\multicolumn{1}{c|}{Mixing \,\,\,\,\,\, }&\multicolumn{1}{c}{ }&\multicolumn{1}{c}{ High }&Mixng\,\,\,\,\,\,\\
\hline
\multicolumn{1}{|c|}{ Parameter } & \multicolumn{1}{c|}{ $o$=10 } & \multicolumn{1}{c|}{ $o$=30 }& \multicolumn{1}{c|}{$o$=50} &\multicolumn{1}{c|}{$o$= 10} &\multicolumn{1}{c|}{$o$= 30} & \multicolumn{1}{c|}{$o$=50} \\
\hline

\,\, $b_1$\,\, & \,\, 0.999\,\, &1.001\,\, & \,\, 0.991 & 0.996\,\,&\,\,1.002\,\,&\,\,0.999\,\,\,\,
\\
\,\, $b_2$\,\,&\,\, 2.78\,\, & \,\,5.78\,\,& \,\, 10.72\,\, &\,\, 4.37\,\,& \,\,10.23\,\,&\,\, 15.75\,\,
\\

\,\, $\nu$\,\, \,\, \,\, &\,\, 1.28\,\, & \,\,1.31\,\,& \,\, 1.37\,\, &\,\,1.24\,\,& \,\,1.31\,\,&\,\, 1.33\,\,
\\
\hline
\end{tabular}
\end{center}
\caption{For the fixed value of $N=200$, we calculate the entropy of mixed state for both equal and high Mixing with number of excitation $o$=10,30 and 50.}
\label{tab:high}
\end{table}
\end{center}
The parameter $b_1$ is approximately unity for the ratio of mixed state with ground state. From the numerical results we can interpret  that the parameter $b_2$ increases with $o$ (fixed $c_1$) and decrease with $c_1$ (fixed o) while the parameter $\nu$ decrease with $o$ (fixed $c_1$) and increase with $c_1$ (fixed o),  where $c_1$ is the relative weight of the mixing of excited state and mixed state. The correction term in equation (\ref{MS}) is regarded as a power law term. The behaviour of this term vanishes rapidly for large radial distances (large horizon area) and becomes significant at small radial distances (small horizon area) as shown in figure (\ref{figratio}).

In order to study the behaviour of entropy of the small and large horizon limit (\ref{figratio}), we substitutes the value of ground state entropy $(S=0.294\,(\frac{r_+}{a}))$ in the equation (\ref{MS}). The expression of entropy becomes,

\be
S_{MS}=b_3A\Big(1+\frac{b_4}{A^{\lambda}}\Big).
\label{e2}
\ee
where $A$ is the area of the horizon {\bf $A=2\pi \,r_+$,~~$b_3=0.294\,b_1$,~~$0.294b_2=b_4$ } and $\lambda=(\nu-1)$. The second term in the expression (\ref{e2}) is a correction to the area law resulting from the entanglement. The correction term vanishes at the large horizon limit (due to the negative exponent of $A$) and the area-law is recovered.
 
\begin{figure}[h]
\centering
\includegraphics[width=1.05\textwidth]{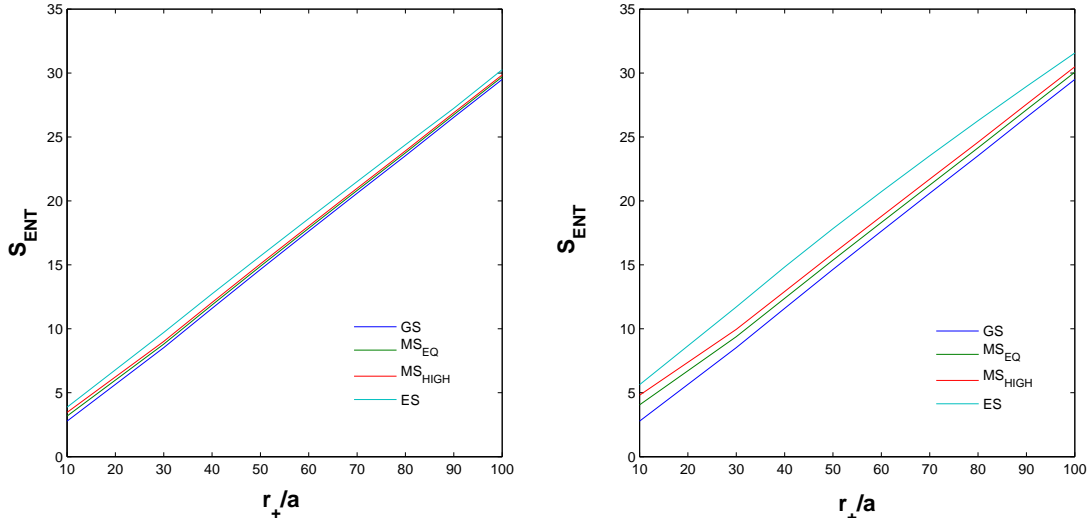}
\caption{ The Entanglement entropy of ground state, excited state and mixed state (Equal and high mixing) are shown as a functions of $r_+/a$ for $N=200$ with excitation number $o$=10 (fig:1) and $o=50$ (fig:2).}
\label{figcomp}
\end{figure}

We compare the ground state, first excited state and mixed states and plot entropy in figure (\ref{figcomp}) with the different value of $o=10$ and $o=50$. In this graph, for the mixed state (equal and high mixing) and excited state, the fitting is nearly linear for the different values of excitation numbers $o=10$ and $o=50$. The entropy of excited and mixed state, coincide with the ground state entropy in the large horizon limit.   

\section{Results and Conclusions}
\label{RES}

In this paper, we have obtained the power law correction to entanglement entropy for both excited state and the mixed state for the scalar field in BTZ black hole spacetime. The ground state follows the Bekenstein-Hawking area law, while the excited state entropy turned out as power of area and the power of area is less then unity. The numerical results show that the area law is violated for the excited state and the power of area deceases with the large area radius as well as excitation numbers. The entropy of mixed states is a linear combination of ground state entropy and the entropy of excited states. The area law is violated not only for the excited state, but also for the superposition of the ground and excited states. For the small horizon area the exponent of area law decreases with increasing the excitation parameters. Thus, the area law exhibits dependence upon the choice of states. 

Here we have investigated the power law corrections using entanglement entropy approach. These corrections to area law have their origin in excited degrees of freedom near the horizon and are significant only for small black hole. In the large horizon limit, the entanglement entropy behaves like thermodynamic entropy and the leading area term describe the situation in which thermal fluctuations are averaged out. However, these {\it thermal fluctuations} are important in case of small black holes and the excited states also contribute to entropy in the form of power law corrections. Also, there are subleading logarithmic corrections to entropy which arise due to the high-energy (short distance) {\it quantum fluctuations} of the fields near the horizon. However, these quantum corrections are small for macroscopic black holes and leading area term describes the thermodynamic entropy of system. 

It would be interesting to investigate the corrections to the black hole entropy with higher curvature terms and higher excited states. We can also extend our results for the fermion fields in the BTZ black hole background \cite{DS4}. 
 
\section*{Acknowledgements}
I would like to thank Dr. Sanjay Siwach for useful discussion. This work is supported by the Rajiv Gandhi National Fellowship Scheme of University Grant Commission (Under the fellowship award no. F.14-2(SC)/2008 (SA-III)), Government of India.

\end {document}